\documentclass[twocolumn,tighten]{aastex7}

\usepackage{amsmath}
\usepackage{subcaption}

\begin{document}

\title{Binary stars take what they get: \\
Evidence for Efficient Mass Transfer from Stripped Stars with Rapidly Rotating Companions}

\author[orcid=0000-0001-6284-2842]{Thibault Lechien}
\affiliation{Max Planck Institute for Astrophysics, Karl-Schwarzschild-Straße 1, 85748 Garching bei München, Germany}
\email[show]{lechien@mpa-garching.mpg.de}
\correspondingauthor{Thibault Lechien}

\author[orcid=0000-0001-9336-2825]{Selma E. de Mink}
\affiliation{Max Planck Institute for Astrophysics, Karl-Schwarzschild-Straße 1, 85748 Garching bei München, Germany}
\email{SEdeMink@mpa-garching.mpg.de}

\author[orcid=0000-0003-3456-3349]{Ruggero Valli}
\affiliation{Max Planck Institute for Astrophysics, Karl-Schwarzschild-Straße 1, 85748 Garching bei München, Germany}
\email{ruvalli@mpa-garching.mpg.de}

\author[orcid=0000-0002-2490-1562]{Amanda C. Rubio}
\affiliation{Max Planck Institute for Astrophysics, Karl-Schwarzschild-Straße 1, 85748 Garching bei München, Germany}
\affiliation{Instituto de Astronomia, Geofísica e Ciências Atmosféricas, Universidade de São Paulo, \\ Rua do Matão 1226, Cidade Universitária, 05508-900 São Paulo, SP, Brazil}
\affiliation{European Organisation for Astronomical Research in the Southern Hemisphere (ESO),  \\ Karl-Schwarzschild-Straße 2, 85748 Garching bei München, Germany}
\email{rubio@mpa-garching.mpg.de}

\author[orcid=0000-0001-5484-4987]{Lieke A. C. van Son}
\affiliation{Center for Computational Astrophysics, Flatiron Institute, 162 Fifth Avenue, New York, NY 10010, USA}
\affiliation{Department of Astrophysical Sciences, Princeton University, 4 Ivy Lane, Princeton, NJ 08544, USA}
\email{lvanson@flatironinstitute.org}

\author[orcid=0000-0002-4313-0169]{Robert Klement}
\affiliation{European Organisation for Astronomical Research in the Southern Hemisphere (ESO), Casilla 19001, Santiago 19, Chile}
\email{robertklement@gmail.com}

\author[orcid=0000-0003-2946-9390]{Harim Jin}
\affiliation{Argelander Institut für Astronomie, Auf dem Hügel 71, DE-53121 Bonn, Germany}
\email{hjin@astro.uni-bonn.de}

\author[]{Onno Pols}
\affiliation{Department of Astrophysics/IMAPP, Radboud University Nijmegen, P.O. Box 9010, 6500 GL Nijmegen, The Netherlands}
\email{o.pols@astro.ru.nl}

\begin{abstract}
Binary stars and their interactions shape the formation of compact binaries, supernovae, and gravitational wave sources.
The efficiency of mass transfer - the fraction of mass retained by the accretor during binary interaction - is a critical parameter that significantly impacts the final fate of these systems.
However, this parameter is observationally poorly constrained due to a scarcity of well-characterized post-mass-transfer binaries.
Be+sdOB binaries, consisting of a rapidly rotating Be star and a stripped hot subdwarf companion, are particularly valuable for studying mass transfer since they represent clear examples of past binary interaction.
Recently, a significantly expanded observational sample of 16 Be+sdOB binaries with well-constrained masses was obtained through combined spectroscopic and interferometric observations.
In this work, we compile and analyze this sample to provide robust constraints on the mass transfer efficiency in binaries that underwent stable mass transfer during the donor's hydrogen-shell burning phase.
Our analysis reveals that mass transfer was predominantly conservative: half of the systems require mass transfer efficiencies above 50\%.
This challenges commonly adopted assumptions of highly non-conservative mass transfer in binary evolution modeling.
Our findings are inconsistent with models that account for spin-up and limit accretion due to a centrifugal barrier.
We also find tension with a commonly used mass transfer model in rapid population synthesis that limits accretion based on the thermal timescale of the accretor.
These results have strong implications for almost all products of binary evolution including the variety of supernovae, white dwarfs, blue stragglers, runaway stars, X-ray binaries, and gravitational-wave sources.

\end{abstract}


\section{Introduction}\label{s: Introduction}

Binary stars are fundamental building blocks of stellar populations,
with many systems orbiting close enough to interact during their lifetime \citep{sana_binary_2012,offner_origin_2023}.
Interacting binaries give rise to a diverse array of phenomena, including blue stragglers \citep{knigge_binary_2009}, %
Type Ia supernovae \citep{whelan_binaries_1973},
X-ray binaries \citep{van_den_heuvel_nature_1973},
stripped-envelope supernovae
\citep{podsiadlowski_presupernova_1992},
and gravitational wave events
\citep{abbott_astrophysical_2016}.

A major uncertainty in our understanding of binary interaction is
the mass transfer efficiency - the fraction of mass retained by the accretor during phases of mass transfer.
This critically influences the final masses and orbital properties, ultimately determining their evolutionary fate.
Theoretical understanding remains poor due to the complex interplay of physical processes involved.
These include the response of the donor and the Roche lobe \citep{soberman_stability_1997}, the accretor's response to gaining mass \citep{kippenhahn_radii_1977,neo_effect_1977, schurmann_exploring_2024} and angular momentum \citep{packet_spin-up_1981, wellstein_prasupernovaentwicklung_2001, langer_binary_2003}, and the interaction of the accretion stream with the companion or the accretion disk \citep{van_rensbergen_spin-up_2008}.

There are two commonly used, physically inspired approaches to model mass transfer.
One approach limits accretion once the mass gainer reaches critical rotation after gaining angular momentum (AM) from the transferred mass.
This \textit{rotationally limited} model predicts that only a small percentage of the mass can be accreted, except in close binaries where tides can prevent spin-up \citep{packet_spin-up_1981,langer_binary_2003,petrovic_constraining_2005,ghodla_evaluating_2023}.
This formalism has been implemented in the widely used MESA evolutionary code \citep{paxton_modules_2015} and is now used in many detailed binary evolution grids, although accretion from a disk is not necessarily expected to stop upon reaching critical rotation, as argued by \citet{paczynski_polytropic_1991}.
The second approach is a \textit{thermally limited} model, which limits accretion if the thermal timescales of both stars differ strongly at the time of mass transfer. A common implementation limits accretion to 10 times the thermal rate of the accretor, aiming to allow for expansion while avoiding a contact situation \citep{tout_rapid_1997,hurley_evolution_2002}. This model allows for a wider range of possible mass transfer efficiencies depending on the size, luminosity, and mass of the stars.
A simpler alternative uses a fixed mass transfer efficiency.
Both this and the thermally limited approach are common in rapid population synthesis models.

Be+sdOB binaries are particularly valuable for studying mass transfer, as they represent clear examples of past binary interaction \citep{pols_formation_1991}.
Hot subdwarfs (sdOB) are stripped, compact stars that have lost their outer hydrogen envelope, leaving behind a helium core.  Be stars are rapidly rotating stars that show emission lines in their spectra due to circumstellar material in a decretion disk \citep{rivinius_classical_2013,rivinius_classical_2024}.
Their fast rotation can be caused by a high rotation rate at birth \citep{bodenheimer_rapidly_1971}, by AM transport from the core to the envelope during main sequence evolution
\citep{ekstrom_evolution_2008, georgy_populations_2013,hastings_single_2020}, or by binary mass transfer \citep{pols_formation_1991}.
Among these, binary interaction appears particularly important, as suggested by population studies \citep[][see however \citealt{van_bever_number_1997,hastings_stringent_2021}]{de_mink_rotation_2013, shao_formation_2014,dodd_gaia_2024,bodensteiner_binarity_2025} and the growing number of confirmed Be binaries with stripped companions \citep[e.g.][]{wang_detection_2021, klement_chara_2024}.

The prototypical Be+sdOB system $\phi$ Persei
provided crucial insights into the evolution of such systems.
The inferred mass transfer efficiency was highly conservative, with the Be star in $\phi$ Persei having accreted at least 70\% of the transferred mass \citep{pols_mass_2007,schootemeijer_clues_2018}.
This result challenged theoretical expectations, specifically those resulting from the rotationally limited mass transfer model. However, with a sample size of one system, it was unclear whether $\phi$ Persei was simply a peculiar case or if it represented a larger class of conservative post-mass transfer binaries. At a lower mass and later evolutionary stage, another well-known example is Regulus, which was also found to have accreted at least 70\% of the transferred mass \citep{rappaport_past_2009}.

Until recently, observations of stripped stars in binaries with main sequence companions have been extremely scarce due to the difficulty in detecting them.
Stripped stars emit mainly in the UV and are easily outshined by their optically bright and typically more massive companions \citep{de_mink_incidence_2014, gotberg_ionizing_2017, gotberg_spectral_2018,schootemeijer_clues_2018}.
Rapid rotation and variability of the companion make it only more challenging. The increased interest in  searching for binaries with black hole companions inadvertently led to the discovery of several stripped star systems
\citep[e.g.\ ][]{
shenar_hidden_2020, %
bodensteiner_is_2020, %
el-badry_stripped-companion_2021, %
el-badry_ngc_2022}, while reanalysis of archival UV datasets led to the discovery of a population in the Magellanic Clouds \citep{drout_observed_2023, gotberg_stellar_2023-1}. For Be+sdOB binaries, the combination of UV spectroscopy and high angular resolution interferometry with the CHARA Array and VLTI/GRAVITY has led to a dramatic increase in the sample size to a total of 16 systems with determined masses for both components \citep{wang_orbital_2023, klement_chara_2024,klement_vltigravity_2025}.

We compile and analyze this new sample of 16 Be+sdOB systems to place lower bounds on the fraction of transferred mass that must have been retained by the accretor. We find evidence for highly conservative mass transfer, with half of the systems requiring that 50\% or more of the transferred mass was accreted.
Our findings are in stark contrast with predictions of models that adopt rotationally limited accretion and, arguably, also in tension with models that use thermally limited accretion (as commonly implemented in rapid population synthesis codes).
This has strong implications for almost all products of binary evolution, due to its effect on the stable mass transfer channel.

\section{Method}\label{s: Methods}
For this study, we compile a sample of Be+sdOB systems with reliable estimates of their masses and orbits.  Most systems in our sample come from the recent work by  \cite{wang_orbital_2023} and \citet{klement_chara_2024, klement_vltigravity_2025}, see Table~\ref{table:sample1} for an overview. Complete dynamical solutions are available for two systems, based on astrometry and radial velocities for both components (which we will refer to as Tier~1 systems, marked with `**').
Seven systems have partial dynamical solutions, meaning that astrometry and radial velocities are only available for the Be star (Tier~2, marked with `*'). Seven further systems only have a spectroscopic characterization (Tier~3). The parameters for the systems in the first two tiers are the most reliable, see Appendix~\ref{sec:appendix_observed_systems}.

Our objective is to constrain the mass transfer efficiency $\beta$, defined as the fraction of mass lost by the donor that is accreted by its companion ($\beta \equiv \Delta M_{\mathrm{acc}}/\Delta M_{\mathrm{donor}}$).
We follow a method adapted from \citet{pols_mass_2007}, which utilizes the fact that the present-day subdwarf star (with mass $M_{\rm sdOB}$) corresponds to the helium core of the donor star. Stellar evolution models link this core mass to the donor's initial mass ($M_{\mathrm{donor,initial}}$). This inferred initial mass allows us to calculate the total mass lost by the donor ($\Delta M_{\mathrm{donor}}$).
At the same time, there is a fundamental evolutionary constraint that the donor must have been initially more massive than the accretor ($M_{\mathrm{donor,initial}} > M_{\mathrm{acc,initial}}$) to evolve and overflow its Roche lobe first.
If the present-day mass of the accretor ($M_{\rm Be}$) is bigger than this upper limit on $M_{\mathrm{acc,initial}}$, that means it must have accreted at least $ \Delta M_{\mathrm{acc}} = M_{\rm Be} - M_{\mathrm{acc,initial}} $, which translates to a lower limit on $\beta$.

\begin{figure}
  \centering
  \includegraphics[trim=6pt 30pt 90pt 40pt, clip,width=\columnwidth]{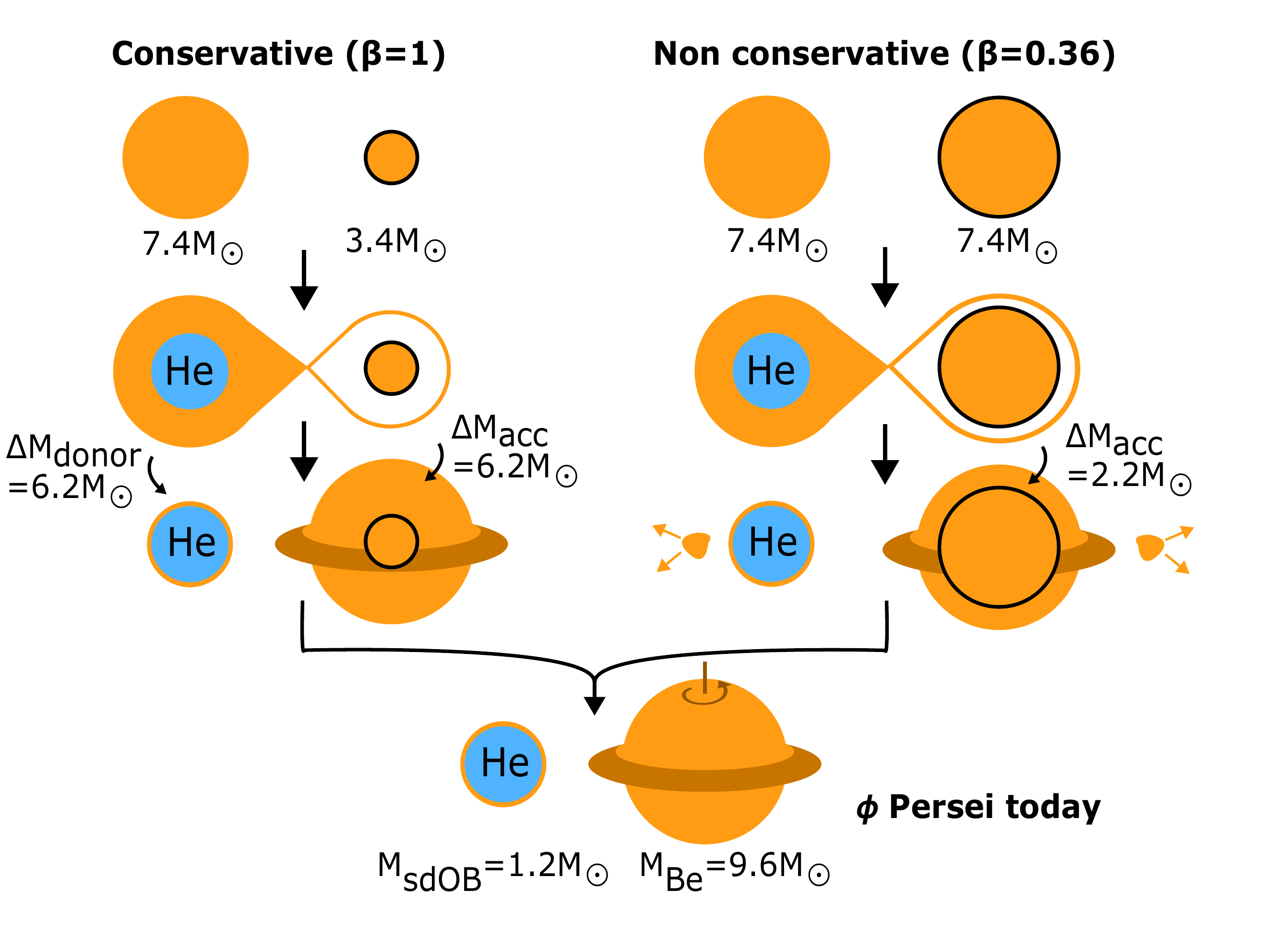}
  \caption{ \textbf{Illustration of the origin of a $\phi$ Persei-like system}, shown for a conservative case (left) and a less conservative case (right).
  The top row shows the ZAMS configuration.
  The second row shows the Roche-lobe overflow, where the donor star fills its Roche lobe and transfers mass to the accretor.
  The third row shows the post-mass transfer configuration, where the donor star has evolved into a stripped helium star, while the accretor has gained mass and AM to become the observed Be star. In the non-conservative case, leftover matter is ejected from the system.
  The bottom row shows the current observed configuration, where the donor is the sdOB star, and the accretor is the Be star.
  }
  \label{fig:combined_cartoon}
\end{figure}

Figure~\ref{fig:combined_cartoon} illustrates this concept using $\phi$ Persei: the left panel depicts a scenario with fully conservative mass transfer ($\beta = 1$), where the accretor gains all the mass lost by the donor.
The right panel shows the case where the efficiency is at its minimum allowed value, $\beta_{\rm min}$, which is where the progenitors had nearly equal initial masses.
Any efficiency lower than $\beta_{\rm min}$ would necessitate $M_{\mathrm{donor,initial}} < M_{\mathrm{acc,initial}}$, violating the evolutionary constraint that the initially more massive star becomes the donor (the sdOB progenitor).

This approach critically depends on an accurate relationship between a star's initial mass and its subsequent stripped (core) mass. We establish this relationship using detailed stellar evolution models computed with MESA (see Appendix~\ref{sec:appendix_mesa_config} for details, including an observationally calibrated, mass-dependent overshooting prescription).
For stars between $2-10$ M$_\odot$ and for the early Case B mass transfer scenario relevant here, we find a tight correlation between the initial mass and the resulting stripped star mass (see Figure~\ref{fig:MinitMstrip}).

To ensure robust constraints, our primary analysis utilizes a relation representing a strict lower limit to the stripped masses found in detailed binary evolution simulations (relation (1) in Figure~\ref{fig:MinitMstrip}), yielding conservative lower bounds on $\beta$.
We also present results using a relation that better reflects the median predicted stripped masses from these simulations (relation (2) in Figure~\ref{fig:MinitMstrip}), which we consider our best-guess estimate.
These relationships allow us to derive analytical expressions for the minimum mass transfer efficiency $\beta_{\rm min}$ as a function of the observed component masses ($M_{\rm sdOB}$ and $M_{\rm Be}$), presented in Equations~\ref{eq:beta_approximation} and \ref{eq:beta_approximation_incomplete_stripping} of Appendix~\ref{sec:appendix_mesa_config}.
We also set a conservative upper-bound on $\beta$ by requiring the initial mass ratio $q = M_{\mathrm{acc,inititial}}/M_{\mathrm{donor,initial}} \ge 0.25$. This allows us to account for the fact that systems with more extreme mass ratios are expected to lead to unstable mass transfer \citep{soberman_stability_1997, temmink_coping_2023}. When quoting our best guess for $\beta$, we restrict the initial mass ratios further to $q \ge 0.5$ to account for the fact that systems with more extreme mass ratios likely get in contact due to thermal expansion of the accretor \citep{neo_effect_1977, wellstein_prasupernovaentwicklung_2001, de_mink_efficiency_2007}.

\section{Results}\label{s: Results}
\begin{figure}
  \centering
  \includegraphics[trim=12pt 2pt 10pt 6pt, clip,width=\columnwidth]{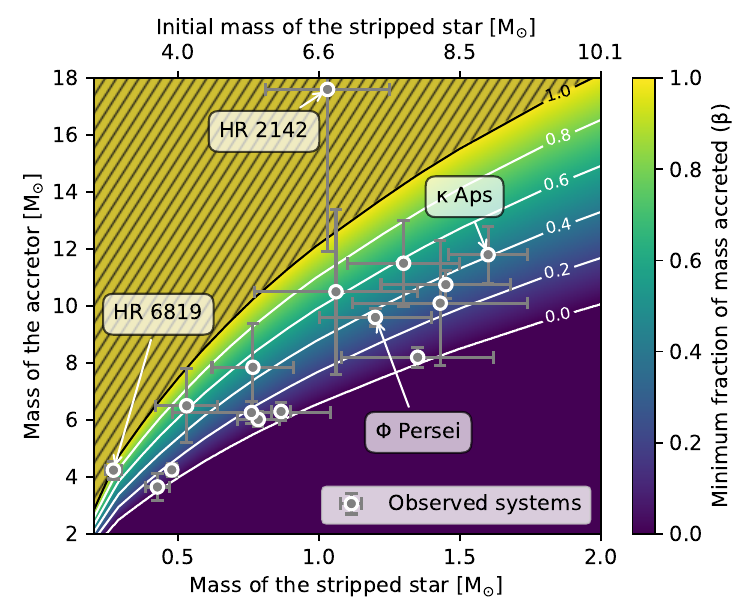}
  \caption{ {\bf Overview of our sample of 16 Be+sdOB binaries.} We plot the present-day mass of the Be star (the rapidly spinning accretor) against that of the sdOB star (the stripped star). We also mark the estimated initial mass using Equation~\ref{eq:init-strip_conservative} on the top axis. The background color gradient and contours show the minimum mass transfer efficiency required to explain the present-day masses of each system, using our conservative estimate (Equation~\ref{eq:beta_approximation}). The hashed region is forbidden under our assumptions as it requires efficiencies above 100\%.   The lower-right side of the diagram is notably empty, even though these systems should be easier to detect.
  }
  \label{fig:MstripMbe}
\end{figure}

Figure~\ref{fig:MstripMbe} shows an overview of the 16 systems in our sample, showing the observed masses of the sdOB and Be stars and their error bars (see Table \ref{table:sample1}).  The color gradient and contours indicate the minimum mass transfer efficiency $\beta_{\rm min}$ required to explain each system, using our strict lower limit estimate.

For two systems, HR 2142 and HR 6819, we strongly expect that they result from Case A mass transfer based on their inferred initial periods (see Appendix~\ref{sec:appendix_beta_constraints}).
Excluding these systems, we find %
that mass transfer was predominantly conservative:
\begin{itemize}
    \item More than a quarter of the sample (4 of 14) requires large accretion efficiencies ($\beta_{\mathrm{min}} > 0.5$).
  \item The majority of the sample (8 of 14) requires moderate accretion ($\beta_{\mathrm{min}} \gtrapprox 0.3$) to explain the median values of the observables.
    \item Even when pessimistically considering the 1-sigma uncertainties, we still find at least 4 systems with evidence for modest accretion ($\beta_{\mathrm{min}} >  0.15$)
\end{itemize}
When using our best guess estimates, the minimum efficiencies are even larger than these strict lower limits.

We note, as can be seen in Figure~\ref{fig:MstripMbe}, that our sample probes systems with inferred initial masses for the subdwarf in the range of $2-9\,{\rm M}_{\odot}$.
The upper left corner of this diagram is empty except for one system with large error bars, which is discussed in detail in Appendix~\ref{sec:appendix_observed_systems}.
This void is expected from both theory (it would require an unphysical mass transfer efficiency larger than 100\%) and observations (systems with extreme mass ratios are hard to detect and characterize due to the large contrast in brightness and limited motion of the Be star).
We further note that the lower right corner is also empty, even though systems here should be easier to detect.  This may be physical and indicate an absence of systems with very low mass transfer efficiencies. It may also be an indication that systems with more extreme initial mass ratios do not evolve through stable mass transfer.
However, since the sample consists of systems collected from the literature, neither the data nor their analysis is homogeneous. Therefore we cannot draw hard conclusions with respect to observational biases.

\begin{figure}[]
  \centering
  \includegraphics[trim=12pt 10pt 10pt 5pt, clip,width=\columnwidth]{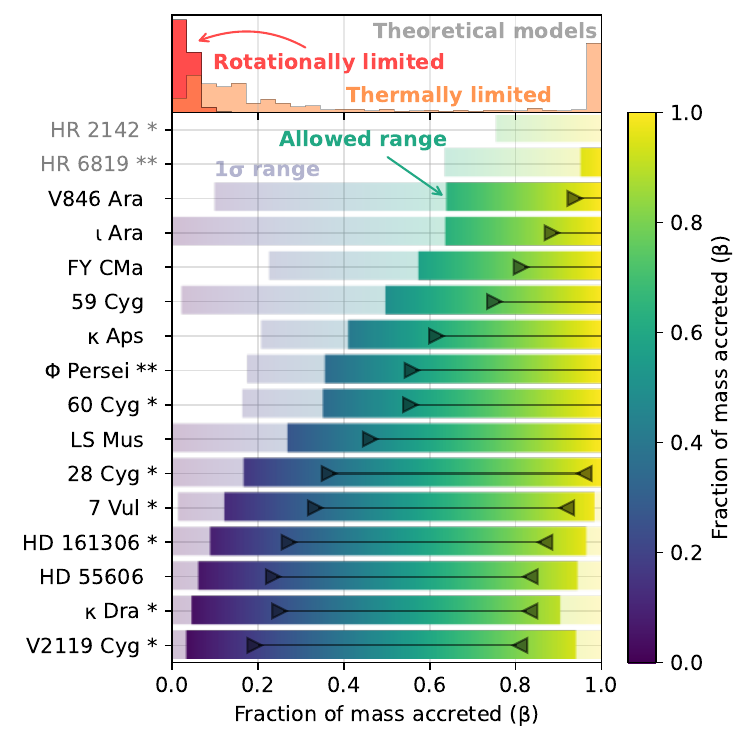}
  \caption{
{\bf The inferred fractions of mass accreted for Be+sdOB binaries challenge theoretical predictions.}
  The color bars in the main panel indicate the allowed ranges for $\beta$ using our strict, conservative assumptions.
  We use the median values for the observed masses, but also show how the range increases when allowing for 1$\sigma$ observational uncertainties.
  Black lines with triangles indicate our best guess range,
  which uses a lower limit based on the median prediction of binary models and an upper limit to exclude systems that likely get into contact (see Section~\ref{s: Methods}).
  The histograms in the top panel show theoretical predictions for early Case B mass transfer, adopting the rotationally limited approach (red, which is not consistent with most systems in our sample) and the thermally limited approach (orange, which is not ruled out here, see however Figure~\ref{fig:schneider_incomplete_10}).
  The first two systems marked in gray are likely the result of Case A mass transfer and cannot be directly compared with the predictions.
}
  \label{fig:allowedRanges}
\end{figure}

\paragraph {Comparison with theory}
The results for $\beta_{\rm min}$ are also shown as color bars in  Figure~\ref{fig:allowedRanges}.
The simplest approach to modeling the mass transfer efficiency, which is still frequently adopted in some rapid population synthesis simulations, is to consider $\beta$ as a parameter and adopt a constant value.
We find that a constant accretion efficiency, where all secondaries accreted 60 -- 80\% of the transferred mass, could in principle explain the entire sample.

Our findings contradict the expectations of the more physically motivated, rotationally limited accretion, which typically allows only a few percent to be accreted  \citep{packet_spin-up_1981, ghodla_evaluating_2023}. To illustrate this, we show theoretical predictions of detailed binary models as a histogram at the top of Figure~\ref{fig:allowedRanges}.
Here we have used the publicly available \texttt{POSYDON} models \citep{fragos_posydon_2023,andrews_posydon_2024} limited to early Case B mass transfer with primaries below 10\,M$_\odot$, but the result is representative of calculations adopting this scheme.
These include many studies using MESA, where this is provided as an option \citep{paxton_modules_2015} based on earlier work by e.g. \cite{wellstein_prasupernovaentwicklung_2001,langer_binary_2003} and \cite{petrovic_constraining_2005}, and other large grids that adopt this assumption (e.g. \citealt{marchant_role_2021, wang_effects_2020,wang_stellar_2022,sen_detailed_2022, xu_populations_2025}).

\begin{figure*}
  \centering
  \includegraphics[trim=10pt 12pt 10pt 10pt, clip,width=0.75\textwidth]{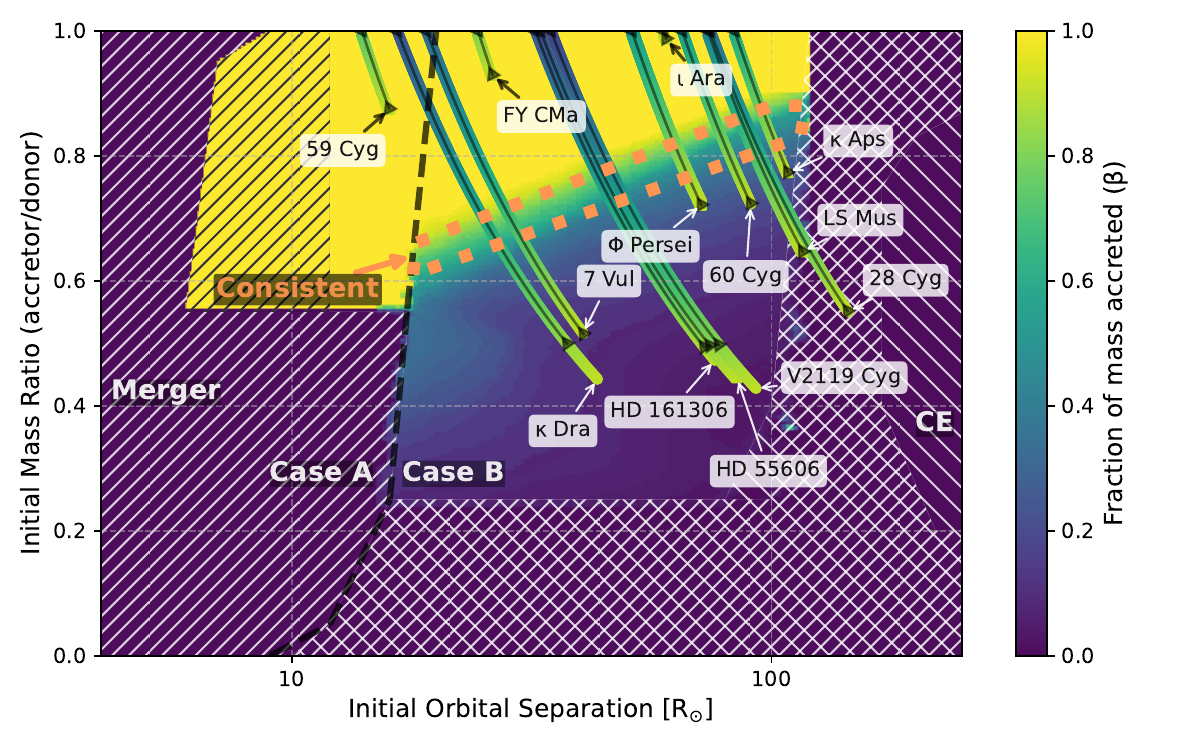}
  \caption{ \textbf{The systems in our sample are also in tension with predictions from the thermally limited model, as implemented in rapid population synthesis simulations.}
  In the background, we show predictions for the fraction of accreted mass ($\beta$, color shading) as a function of the initial mass ratio and orbital separation.
  We overplot our best guess for the initial parameters for each observed system, which depend on the assumed $\beta$ (indicated in color).
  Higher assumed values of $\beta$ for the observed systems imply lower initial mass ratios, which is where this model predicts low $\beta$ values and vice versa.
  The main region in this plot where the data and model predictions are consistent is marked in orange. We deem it unlikely that the vast majority of systems originate from such a narrow region and interpret this as tension with this model (here we have taken  \citealt{schneider_evolution_2015}, their Figure 18, assuming a primary mass of 5 M$_\odot$ as an illustration).}
  \label{fig:schneider_incomplete_10}
\end{figure*}

We also compare with predictions from a model that adopts thermally limited accretion, taken from rapid population synthesis simulations by \citet{schneider_evolution_2015} conducted with \texttt{binary\_c} \citep{izzard_new_2004,izzard_population_2006,izzard_population_2009}, limited here to early Case B mass transfer models for a 5 M$_\odot$ primary star.
This code is based on the original \texttt{BSE} code \citep{hurley_comprehensive_2000,hurley_evolution_2002}. It limits accretion during stable mass transfer to ten times the thermal rate. The thermal timescales are estimated using the \citet{hurley_comprehensive_2000} formulae for the stellar parameters.
This model effectively predicts a bimodal distribution, where the overwhelming majority of systems are either non-conservative ($\beta < 0.2$) or purely conservative ($\beta = 1$). However, our best guess range for roughly half of the sample lies in the intermediate range of $\beta \in (0.2,0.9)$, which does not have much support from these predictions.

Further tension becomes clear in Figure~\ref{fig:schneider_incomplete_10}, where we show the predictions of this thermally limited model as a function of the initial orbital separation and mass ratio.  We overplot our best guess for the inferred initial parameters for each system in our sample. These appear as lines, as they depend on the assumed mass transfer efficiency.  For angular momentum loss, we have assumed here that 10\% of the transferred mass is sent to a circumbinary disk, and the rest is emitted isotropically from the vicinity of the accretor; see Appendix~\ref{sec:appendix_thermally_limited} for alternatives.

For ten systems, we find consistent solutions in the initial parameter space where the mass transfer efficiency predicted by the model matches the mass transfer assumed.
However, all of these lie in a narrow region marked in orange. This is where the thermally limited model undergoes a steep transition from predicting fully conservative accretion to highly non-conservative accretion. This narrow band accounts for approximately 10\% of the initial parameter space for stable mass transfer.  The probability that ten systems in our sample come from this region by chance is extremely small, of the order of $10^{-10}$.  Varying our assumptions for the degree of stripping and angular momentum loss leads to very similar findings.
We conclude that, under our assumptions, this implementation of thermally limited mass transfer is only consistent with a subset of the observed systems and only if the initial parameters are fine-tuned.  We deem such fine-tuning unlikely and conclude that the data indicate a tension with this model.

Although we have only compared with the implementation in \citet{schneider_evolution_2015}, we expect that this tension holds for all rapid population synthesis codes rooted in the original \texttt{BSE} code that adopt the thermally limited scheme.
The thermally limited approach is also adopted in some detailed binary calculations, e.g. \texttt{BPASS} \citep{eldridge_effect_2008,byrne_dependence_2022}.
Detailed calculations, solving for the structural changes of the stars, have the advantage of more accurately following the radial expansion of the accretor as it is driven out of thermal equilibrium.
The increase in radius leads to a decrease of the timescale, implying that it can accrete at a higher rate.
In wider binary systems, significant expansion of the accretor is possible and this should allow for more conservative accretion than the predictions in rapid population synthesis \citep{kippenhahn_radii_1977,lau_expansion_2024}.
This may increase the region in the parameter space where intermediate mass transfer efficiencies are predicted, and thus lower the tension we noted above. However, we cannot state this with certainty since we have not compared this directly.

\paragraph{Correlations with initial parameters}
We also search for possible correlations between our conservative lower bounds on $\beta$, the observed properties of the systems, and their inferred initial states.  We find a strong correlation that the systems with a higher lower bound on $\beta$ are also the ones with inferred initial mass ratios closer to unity.
This fits qualitatively with the expectations of the thermally limited accretion scheme as implemented in rapid population synthesis.
We also find a moderate but still significant correlation between the mass transfer efficiency and the initial period, where initially closer systems are more conservative.
Qualitatively, this again fits the expectations of the thermally limited accretion scheme, because in closer binaries the donor is less expanded and will donate mass on a slower timescale, but also the rotationally limited approach, where tides in close binaries are more efficient in preventing spin-up, hence allowing for more accretion \citep{zahn_dynamical_1975,zahn_tidal_1977,hut_tidal_1981,langer_binary_2003}.
However, we cannot infer the causality of these correlations because of the close ties between these properties and the method by which we infer the minimum mass transfer efficiency.
We do not find a correlation of $\beta$ with the initial donor mass in the mass range we probe of $2 - 9$ M$_\odot$.
The correlations with these properties, in addition to others, can be found in Figure~\ref{fig:correlation} of Appendix~\ref{sec:appendix_beta_constraints}.

\section{Discussion}\label{s: Discussion}
Our analysis reveals strong evidence for highly conservative mass transfer in Be+sdOB binary systems, with many systems requiring that more than 50\% of the transferred mass was retained by the accretor. This finding challenges common assumptions in binary modeling and raises important questions about the physical mechanisms that enable such efficient mass transfer.

To address these questions, several physical processes have been proposed to facilitate efficient accretion.
Tidal interactions can help spin down the accretor, especially in short-period systems \citep{petrovic_constraining_2005,petrovic_which_2005,sen_detailed_2022}. However, tides seem to be insufficient to explain the rotation rate of longer-period systems (P $>$ 5 days, \citealt{dervisoglu_spin_2010}).
A strong large-scale magnetic fields has been detected in at least one star that recently gained mass in a binary system \citep{grunhut_discovery_2013}.  This suggests that spin-down via magnetic braking could be important and has been explored for low-mass stars in binaries \citep{dervisoglu_spin_2010,sun_stellar_2024}.  However, large-scale magnetic fields are not detected in Be stars \citep{wade_mimes_2016}.
The most promising explanation, which has been shown in numerous accretion models of rapidly rotating stars, is that the viscous accretion disk carries away AM efficiently, while still accreting matter on to the star \citep{paczynski_polytropic_1991, popham_does_1991, colpi_analytical_1991,bisnovatyi-kogan_self-consistent_1993,porter_classical_2003,carciofi_nonlte_2008,krticka_mass_2011,deschamps_critically-rotating_2013}.

Many earlier studies have attempted to constrain the mass transfer efficiency with varying findings, which may reflect a dependency on the system parameters.
Our work probes progenitors in the $2-9$ M$_\odot$ range and holds for systems that undergo stable mass transfer where the donor has left the main sequence (early Case B). \cite{bao_be_2025} analyzed a similar sample of Be+He stars using population synthesis models and found that a higher mass transfer efficiency ($\gtrsim 0.5$) better supports the observations, confirming our findings of predominantly conservative mass transfer.
\cite{vinciguerra_be_2020} used constraints on the mass distribution and orbital periods of Be X-ray binaries, which, prior to the supernovae, are thought to evolve through the same channel, albeit at slightly higher masses $>10$ M$_\odot$.  They found that mass transfer efficiencies of at least 30\% are needed to reproduce their observed properties, consistent with our findings.
At lower initial masses of around 1.2~M$\odot$, blue straggler systems can also require high efficiencies, inferring $\beta$ values exceeding 50\% \citep{gosnell_constraining_2019, sun_wocs_2023}, while some point toward an efficiency of around 20\% \citep{sun_wocs_2021}.
In contrast, studies of massive Wolf-Rayet binaries (40-60~M$\odot$) favor non-conservative mass transfer, with inferred efficiencies typically below 50\% and often as low as 10\% \citep{petrovic_constraining_2005, shao_nonconservative_2016,schurmann_exploring_2024, nuijten_wr_2024}.
In addition to using post-mass transfer systems, many authors have derived constraints from systems where mass transfer is ongoing, for example, in semi-detached Algols.
Here, a wide range of efficiencies are required to explain the observed populations, suggesting different efficiencies even within similar systems \citep{de_greve_origin_1994,nelson_complete_2001,de_mink_efficiency_2007,van_rensbergen_spin-up_2008,van_rensbergen_mass_2010,van_rensbergen_mass_2011,mennickent_accretion_2013, mennekens_comparison_2017,dervisoglu_evidence_2018, sen_detailed_2022}.

Some important caveats should be considered when interpreting our results.
We assume an initial mass -- stripped mass relation most appropriate for Case B mass transfer.
However, two systems with relatively short initial periods and high accretion efficiency constraints, namely HR 2142 and HR 6819, have likely evolved through Case A mass transfer.
Case A would strip the star already during its main sequence evolution before the helium core is fully developed. This leads to a lower mass for the stripped star, translating to a larger amount of donated mass and a lower mass transfer efficiency constraint.

Alternatively, a separate spectral fitting of HR 2142 using \textsc{BeAtlas} \citep{rubio_bayesian_2023} found a lower mass for the Be star of 9.6 M$_\odot$.
Combining this information with the orbital parameters presented in \cite{klement_chara_2024}, the mass of the sdOB companion is around 0.7 M$_\odot$, which matches previous estimates \citep{peters_hot_2016}.
This leads to a lower mass transfer efficiency constraint of $\beta_{\mathrm{min}}$ = 0.85, and its initial period could be marginally consistent with Case B.

The initial mass -- stripped mass relation also depends on the treatment of overshooting in stellar models.
The choice in overshooting, treatment above the convective core, and metallicity can affect the core sizes and the mass left after stripping \citep{yoon_type_2017, gotberg_ionizing_2017,laplace_expansion_2020}, although
the stars in our sample are located close enough within the Milky Way that the metallicities are close to solar.

Finally, we did not consider the possible effect of interaction with a third companion in altering the present-day masses of the system. This should be rare enough that it does not affect our overall conclusion, but we cannot rule out that it could have affected an individual system in our sample \citep{offner_origin_2023}.

\section{Conclusion}\label{s: Conclusion}
We derived observational constraints on the mass transfer efficiency by compiling and analyzing a recent, significantly expanded sample of 16 well-characterized Be+sdOB binary systems.
The sample is representative for systems with initial primary masses in the range of $2 - 9$  M$_\odot$  that undergo stable mass transfer after the primary leaves the main sequence (early Case B).
Our key findings are:
\begin{enumerate}
  \item Mass transfer in these systems was predominantly efficient, with half of the systems requiring at least 50\% of the transferred mass to be retained by the accretor, using our best-guess estimates.
  \item The rotationally limited accretion scheme (as implemented in e.g. MESA) is inconsistent with our inferred lower bounds.
  \item There is tension with the thermally limited accretion scheme that is currently implemented in several rapid binary population synthesis codes, as it requires fine-tuned initial parameters for the systems in our sample.
  \item Models that assume a fixed accretion efficiency for stable Case B mass transfer would be able to explain our entire sample if 60 -- 80\% of the transferred mass is accreted.
\end{enumerate}

These results challenge commonly made assumptions in state-of-the-art binary simulations.
Our findings have a large impact on the lives and final fate of binary systems that evolve through a phase of stable mass transfer, as it affects the final masses and orbital configurations.
This in turn influences our understanding of binaries containing white dwarfs, type Ia supernovae, and the properties of blue stragglers, runaway stars, Be stars, and mass gainers in general.
If our findings also apply to somewhat higher mass systems, as supported by earlier work on Be X-ray binary systems, this would have large implications for the variety of core-collapse supernovae, X-ray binaries, and gravitational wave sources.

\section*{Acknowledgements}
We would like to thank the anonymous referee for the
useful feedback.
We also thank Norbert Langer and Emmanouil Zapartas for helpful discussions.
This research has made use of the Astrophysics Data System, funded by NASA under Cooperative Agreement 80NSSC21M00561.

\section*{Software and Data}
This work made use of the following software packages: \texttt{python} \citep{van_rossum_python_2009}, \texttt{numpy} \citep{harris_array_2020}, \texttt{matplotlib} \citep{hunter_matplotlib_2007}, \texttt{scipy} \citep{virtanen_scipy_2020, gommers_scipyscipy_2025}, and Modules for Experiments in Stellar Astrophysics (\texttt{MESA}) \citep{paxton_modules_2011, paxton_modules_2013, paxton_modules_2015, paxton_modules_2018, paxton_modules_2019, jermyn_modules_2023}.
The MESA inlists and history files used for these calculations are available on Zenodo: \dataset[doi:10.5281/zenodo.16052986]{https://doi.org/10.5281/zenodo.16052986}.

\begin{appendix}

\section{Observed systems}
\label{sec:appendix_observed_systems}

We compile a new, significantly expanded sample of well-characterized Be+sdOB binaries, primarily based on \cite{wang_orbital_2023} and \cite{klement_chara_2024}.
The recent increase in the number of such binaries with well-determined parameters was made possible in part due the CHARA (Center for High Angular Resolution Astronomy) Array \citep{ten_brummelaar_first_2005} and VLTI/GRAVITY \citep{gravity_collaboration_first_2017, eisenhauer_advances_2023} instruments, combined with detailed spectroscopic studies using FEROS \citep{kaufer_commissioning_1999} and CHIRON \citep{tokovinin_chironfiber_2013}.
Far-UV spectroscopy (e.g. with the International Ultraviolet Explorer and the Hubble Space Telescope) has been crucial for measuring radial velocities and detecting the faint sdOB companions through their UV signatures, while interferometry has provided direct constraints on orbital geometries and information about possible circumstellar material. These complementary observational techniques have enabled precise measurements of orbital parameters, stellar properties and circumstellar environments, leading to a large new sample of Be stars with stripped companions.

The properties of our full sample of 16 Be+sdOB binaries are presented in Table~\ref{table:sample1}.
We distinguish three types of system according to how they were characterized.

\begin{table}[]
\caption{ \textbf{Properties of Be+sdOB binary systems in the sample.} The tier indicates the quality of the solutions: ``**'': Tier 1, with complete dynamical solutions, ``*'': Tier 2, with only partial dynamical solutions, and ``-'': Tier 3, with only spectroscopic solutions.
On the right, we also give our inferred minimum mass transfer efficiency ($\beta_{\text{min}}$) and our best-guess mass transfer efficiency range, as detailed in Section~\ref{s: Methods}. HR 2142 and HR 6819 likely evolved through Case A mass transfer, so their $\beta$ estimates are less certain and shown in brackets.
}
  \label{table:sample1}
  \centering
  \begin{tabular}{l@{\hspace{4pt}}c@{\hspace{4pt}}c@{\hspace{4pt}}c@{\hspace{4pt}}c@{\hspace{4pt}}l@{\hspace{4pt}}c@{\hspace{3pt}}|@{\hspace{4pt}}c@{\hspace{10pt}}c@{}}
  \hline\hline
  HD & Star & M$_\mathrm{Be}$ & M$_\mathrm{sdOB}$ & Orbital Period  & Reference & Tier & $\beta_{\text{min}}$ & Best-guess \\
  Number & Name & (M$_\odot$) & (M$_\odot$) & (days) &  &  &  & $\beta$ range \\
  \hline
  58978 & FY CMa & 10--13 & 1.1--1.5  & 37.253 $\pm$ 0.007 & \cite{peters_detection_2008} & - & 0.57 & 0.81 - 1.00\\
  200120 & 59 Cyg & 6.3--9.4 & 0.62--0.91 & 28.1871 $\pm$ 0.0011 & \cite{peters_far-ultraviolet_2013} & -& 0.50 & 0.75 - 1.00 \\
  10516 & $\phi$ Per & 9.6 $\pm$ 0.3 & 1.2 $\pm$ 0.2 & 126.6982 $\pm$ 0.0035 & \cite{mourard_spectral_2015} & ** & 0.36 & 0.56 - 1.00\\
  55606 & - & 6.0--6.6 & 0.83--0.90 &  93.76 $\pm$ 0.02 & \cite{chojnowski_remarkable_2018} & - & 0.06 & 0.23 - 0.83 \\
  109387 & $\kappa$ Dra & 3.65 $\pm$ 0.48 & 0.426 $\pm$ 0.043 & 61.5496 $\pm$ 0.0058 & \cite{klement_dynamical_2022}  & * & 0.05 & 0.25 - 0.83\\
  113120 & LS Mus & 10.1 $\pm$ 2.2 & 1.43 $\pm$ 0.31 & 181.54 $\pm$ 0.11 & \cite{wang_orbital_2023}  & - & 0.27 & 0.46 - 1.00\\
  137387 & $\kappa$ Aps & 11.8 $\pm$ 1.0 & 1.60 $\pm$ 0.14 &  192.1 $\pm$ 0.1 &\cite{wang_orbital_2023}  & -& 0.41  & 0.62 - 1.00 \\
  152478 & V846 Ara & 6.5 $\pm$ 1.3 & 0.53 $\pm$ 0.11 & 236.50 $\pm$ 0.18 & \cite{wang_orbital_2023} & - & 0.64 & 0.94 - 1.00 \\
  157042 & $\iota$ Ara & 10.5 $\pm$ 2.9 & 1.06 $\pm$ 0.29 & 176.17 $\pm$ 0.04 & \cite{wang_orbital_2023} & - & 0.64  & 0.88 - 1.00\\
  41335 & HR 2142 & 17.6 $\pm$ 5.7 & 1.03 $\pm$ 0.22 &  80.8733 $\pm$ 0.0044 &  \cite{klement_chara_2024}  & * & (1.00) & (1.00 - 1.00) \\
  161306 & - & 6.02 $\pm$ 0.26 & 0.784 $\pm$ 0.074 & 99.315 $\pm$ 0.055 & \cite{klement_chara_2024} & * & 0.09 & 0.27 - 0.87 \\
  183537 & 7 Vul & 4.25 $\pm$ 0.23 & 0.477 $\pm$ 0.020 & 69.5258 $\pm$ 0.0051 & \cite{klement_chara_2024}  & * & 0.12 & 0.33 - 0.92 \\
  191610 & 28 Cyg & 6.26 $\pm$ 0.38 & 0.76 $\pm$ 0.28 & 359.260 $\pm$ 0.041 & \cite{klement_chara_2024} & * & 0.17 & 0.37 - 0.96 \\
  194335 & V2119 Cyg & 8.20 $\pm$ 0.35 & 1.35 $\pm$ 0.27 & 63.1475 $\pm$ 0.0029 & \cite{klement_chara_2024} & * & 0.03 & 0.19 - 0.81 \\
  200310 & 60 Cyg & 10.75 $\pm$ 0.49 & 1.45 $\pm$ 0.23 & 147.617 $\pm$ 0.038 & \cite{klement_chara_2024} & * & 0.35 & 0.55 - 1.00  \\
  167128 & HR 6819 & 4.24 $\pm$ 0.31 & 0.270 $\pm$ 0.027 & 40.3261 $\pm$ 0.0013 & \cite{klement_vltigravity_2025}  & ** & (0.95) & (1.00 - 1.00) \\
  \hline
  \end{tabular}
\end{table}

\paragraph{Tier 1 ** (Complete Dynamical Solutions) -- $\phi$ Per and HR\,6819:}
These systems were characterized using both interferometric and spectroscopic observations. For $\phi$ Per, the interferometric data came from CHARA, using the MIRC instrument in H-band (see also \citealt{klement_interferometric_2022}). The spectroscopic analysis utilized archival radial velocity measurements from optical and UV spectra, including observations from the International Ultraviolet Explorer (IUE) and the Hubble Space Telescope (HST). By combining astrometric measurements with radial velocity solutions and assuming a circular orbit, \cite{mourard_spectral_2015} were able to determine both the orbital parameters and stellar masses of the system.
For HR 6819, the orbital solution was derived by combining radial velocity measurements from the spectroscopic data with high-precision astrometric positions from GRAVITY.
It was proposed to be a black hole candidate \citep{rivinius_naked-eye_2020}, but is now considered to be a stripped star system \citep{bodensteiner_is_2020,gies_h_2020,el-badry_stripped-companion_2021,frost_hr_2022}.

For both systems, relative astrometry and RV's for both components were measured. This allowed for the determination of the dynamical masses of both components, in addition to the dynamical parallax, without the need for additional distance measurements.

\paragraph{Tier 2 * (Partial Dynamical Solutions) -- HD\,161306, 7\,Vul, 28\,Cyg, V2119\,Cyg, 60\,Cyg, $\kappa$\,Dra, and HR\,2142: }
High-angular resolution interferometric observations were obtained using the CHARA Array with both MIRC-X and MYSTIC instruments in the near-infrared H-band and K-band respectively. For HR\,2142, additional spectro-interferometric data from VLTI/GRAVITY provided complementary constraints. The analysis also incorporated archival optical spectroscopy from the literature. By combining radial velocity measurements from spectroscopy with interferometric observations, orbital solutions were derived.
The dynamical masses of the components were then determined by incorporating the distances obtained from Gaia measurements.

Note that RV's were only available for one of the two components - either from optical spectroscopy of the Be star or from UV spectroscopy of the sdOB companion. This means that any error in the distance measurement will directly affect the mass determination of that system.

\paragraph{Tier 3 (Spectroscopic Solutions) -- FY\,CMa, 59\,Cyg, HD\,55606, LS\,Mus, $\kappa$\, Aps, V846\,Ara, and $\iota$\,Ara: }
For FY\,CMa and 59\,Cyg, double-lined spectroscopic orbits were derived from UV spectra using cross-correlation and Doppler tomography techniques, with stellar properties determined by fitting separated spectra with photospheric models. HD\,55606 was characterized through spectroscopic observations that revealed double-lined features, allowing for a spectroscopic orbital solution. The remaining systems (LS\,Mus, $\kappa$\, Aps, V846\,Ara, $\iota$\,Ara) were analyzed using far-UV and optical spectroscopy, with sdOB companion masses derived from orbital mass ratios and Be star masses adopted from literature.

These systems lack astrometric or interferometric measurements, resulting in less precise estimates of orbital inclinations.
Therefore, the mass determinations for these systems are less certain than those with direct dynamical measurements.

\paragraph{Further comments on individual systems: }
For three systems (FY CMa, 59 Cyg, and HD 55606), the literature provided ranges rather than specific values with uncertainties. In these cases, we converted the ranges into median values with uncertainties spanning the full range (±1 sigma).
For HR 6819, multiple orbital solutions exist; we adopted the preferred, slightly eccentric solution from \cite{klement_vltigravity_2025}, which combines both astrometric and radial velocity measurements.
We note that alternative solutions for this system do not significantly affect our derived mass transfer efficiency constraints.

Among our sample, 60 Cyg exhibits the highest orbital eccentricity at 0.20 $\pm$ 0.01, followed by 59 Cyg with an eccentricity of 0.14.
This is surprising for post mass transfer systems: most simulations assume that strong tidal interaction would lead to circularization of the orbit.
For 59 Cyg, \cite{klement_chara_2024} was able to detect a third outer component on an eccentric orbit which can affect the inner binary, but this has not been detected for 60 Cyg.
The remaining systems have not been detected to be part of a hierarchical triple or higher-order system, nor have significant eccentricities.

HR 2142 shows an unusually high Be star mass, while previous work \citep{peters_hot_2016} and a separate spectral fitting using \textsc{BeAtlas} \citep{rubio_bayesian_2023} suggest a lower Be star mass ($\approx$ 9 M$_\odot$) than the one derived from orbital parameters (17.6 M$_\odot$).
This indicates possible systematic uncertainties in the mass determination, particularly given that other systems in our sample show better agreement between spectral and orbital mass estimates.
A possible cause could be the lack of interferometric observations or spectroscopic orbital solution of the sdOB star \citep{klement_chara_2024}.
The system is also notable because of the obscuration of the companion by the Be star.

Finally, LS Mus, k Aps, and 60 Cyg have an estimated sdOB mass of above 1.4 M$_\odot$, the most massive in our sample \citep{wang_orbital_2023}.  This implies that they are close to the regime for progenitors of Type Ib or Ic supernovae. A future supernova explosion and formation of a neutron star could then lead to the formation of a Be X-ray binary if the system remains bound.

\section{initial mass -- stripped mass relation}
\label{sec:appendix_mesa_config}
To investigate the relationship between the initial mass and the mass of the stripped star, we employ the detailed one-dimensional stellar evolution code MESA \citep{paxton_modules_2011, paxton_modules_2013, paxton_modules_2015, paxton_modules_2018, paxton_modules_2019, jermyn_modules_2023} version \texttt{r23.05.1}.
This code solves the coupled equations of stellar structure and nuclear burning to model the one-dimensional evolution of stars.
We detail the key physical ingredients of our models below:

\begin{itemize}
\item We adopted a solar metallicity of Z = 0.014 with the default MESA solar mixture.

\item For convective boundaries, we implemented a mass-dependent overshooting prescription where the convective zone is extended by $\alpha_\mathrm{ov}$ pressure scale heights. Following observational constraints \citep{castro_spectroscopic_2014,claret_dependence_2016,schootemeijer_constraining_2019}, $\alpha_\mathrm{ov}$ increases linearly from 0.1 at 1.66 M$_\odot$ to 0.3 at 20 M$_\odot$ \citep{brott_rotating_2011}.

\item We used the nuclear reaction network with 21 isotopes (`approx21.net'), which includes the main reactions relevant for hydrogen and helium burning.

\item We included mass loss through stellar winds using the ``Dutch" wind scheme \citep{glebbeek_evolution_2009}.
However, the effect of wind mass loss is minimal during the main sequence and early post-main sequence evolution relevant for our analysis.

\item We used the Ledoux criterion for convective stability with a semiconvection efficiency parameter of $\alpha_\mathrm{sc}$ = 1.0 \citep{langer_evolution_1985}. %
\end{itemize}

We compute the evolution of 18 single-star, solar metallicity MESA models with Zero-age Main Sequence (ZAMS) masses between 2 and 10.5 M$_\odot$ with 0.5 M$_\odot$ increments.
Based on these models, we derive a relationship between the ZAMS mass and the helium core mass during the hydrogen shell burning phase.
For our analysis, we define the hydrogen shell burning phase as starting when the central hydrogen abundance drops below $10^{-5}$ (signifying the end of core hydrogen burning) and ending when the central carbon abundance rises above $2 \times 10^{-4}$ (signifying the start of core helium burning).
We define the helium core mass as the mass coordinate where the hydrogen mass fraction drops below 0.1 and the helium mass fraction is above 0.1.
We then use linear interpolation between the model values to derive the relationship between the initial mass and helium core mass.

\begin{figure}[]
  \centering
  \includegraphics[trim=10pt 12pt 10pt 10pt, clip,width=.65\columnwidth]{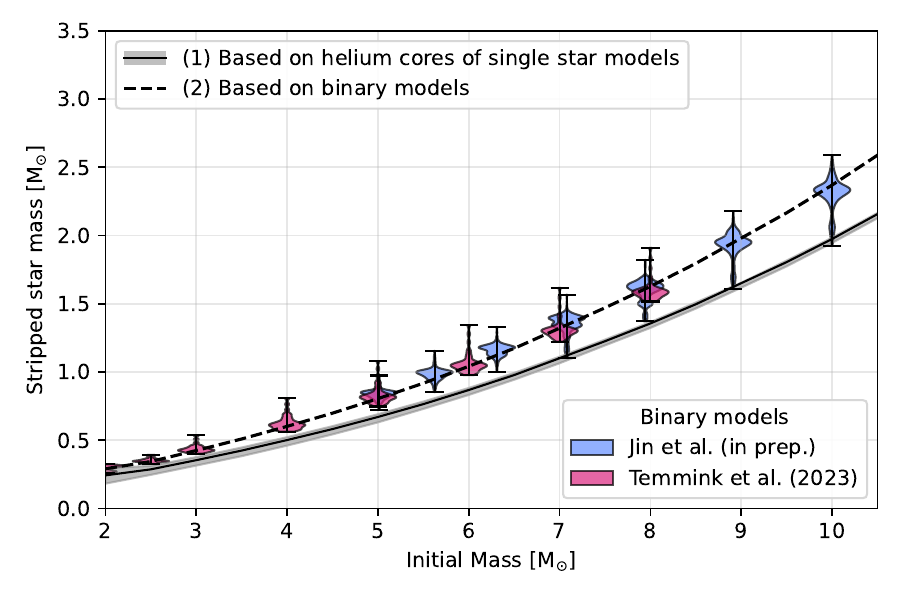}
  \caption{ \textbf{Relationship between the initial mass and stripped star mass.}
  The solid black line (1) shows the helium core mass derived from our grid of single-star MESA models, taken during the hydrogen shell burning phase.
  The light gray background shows the slight variation in helium core mass during this phase.
  The helium core mass is defined as the mass coordinate where the hydrogen mass fraction drops below 0.1 and the helium mass fraction is above 0.1.
  The violin plots show the range of post-stripping masses in two large grids of early Case B binary MESA models (Jin et al. (in prep.) and \cite{temmink_coping_2023}), indicating that the helium core mass is a lower bound for the final mass of the stripped star.
  The dashed line (2) shows the helium core mass relation, with an added 20\% of the mass remaining, which is a good approximation of the incomplete stripping that occurs during Case B mass transfer.
  }
  \label{fig:MinitMstrip}
\end{figure}

The initial mass -- helium core mass relationship is shown in Figure~\ref{fig:MinitMstrip} as a solid black line (1). The gray area around the black line shows the variation depending on the stage of hydrogen shell burning that is chosen.
The helium core mass does not change significantly during this phase, so the value at the middle time of the phase is a good approximation for the core mass, which is what we use for our strict mass transfer efficiency constraints.
An analytical fit to the initial mass -- helium core mass (1) is given as follows and is accurate to within 0.1 M$_\odot$ of the computed models (where the units of the numerical coefficients are such that all masses are in units of M$_\odot$):

\begin{equation}
  M_{\text{init}} = 10.046 \cdot (M_{\text{strip}})^{0.494} - 0.634 \cdot M_{\text{strip}} - 2.820
  \label{eq:init-strip_conservative}
\end{equation}

We compare our relation with the results derived from full binary evolutionary models computed with MESA.
The blue violin plots show the range of masses post-stripping from a grid of early Case B binary models from Jin et al. (in prep.), using the same overshooting prescription as our single-star models. We only include binary models where the donor star has a radiative envelope at the onset of mass transfer.
We also show the remnant masses of early Case B mass transfer from a different grid of MESA models by \cite{temmink_coping_2023} in red violin plots, aggregating over initial mass ratios and orbital separations.
The \cite{temmink_coping_2023} models adopt overshooting as calibrated by \citet{choi_mesa_2016} based on the star cluster M67, which has a turn of mass of about 1.25 M$_\odot$ \citep{sandquist_k2_2021}. This is a good choice for the lower end of our mass range, but may not apply to the high mass end.

The figure shows that the relation we use in our main analysis, the solid black line (1), is a good lower bound for the final mass of the stripped star, as it lies at the lower end of the stripped mass ranges covered by the binary models.
However, as the binary models show, in most cases the mass post-stripping is 10-25\% higher due to the incomplete stripping of the envelope. We also analyze our systems using the relation shown as a dashed line (2), which considers an added 20\% of the mass remaining. It does not provide a strict lower bound, but we consider this our best guess.
An analytical fit to the initial mass -- stripped mass relation (2) is given as follows and is accurate to within 0.1 M$_\odot$ of the computed models:
\begin{equation}
  M_{\text{init}} = 9.335 \cdot (M_{\text{strip}})^{0.457} - 0.263 \cdot M_{\text{strip}} - 3.236
  \label{eq:init-strip_bestguess}
\end{equation}

As the paper detailing the binary grid models is forthcoming (Jin et al. in prep), we briefly summarize their key assumptions here.
They use MESA version 10398 \citep{paxton_modules_2011, paxton_modules_2013, paxton_modules_2015, paxton_modules_2018, paxton_modules_2019} to model the evolution of binary systems with initial primary masses ranging from 5 to 100 M$_\odot$ (in steps of $\Delta \log M_{1,i} = 0.05$ M$_\odot$), initial mass ratios from 0.1 to 0.95 (in steps of $\Delta q_i = 0.05$), and initial orbital periods between 0.3 and 5000 days (in steps of $\Delta \log P_i = 0.05$ days). The individual stellar components are modeled using the same physical assumptions as in \cite{jin_boron_2024}, which crucially uses the same overshooting prescription as our single-star models.
The binary physics follows \cite{marchant_ultra-luminous_2017}, except for a different mass transfer scheme. During core hydrogen burning, mass transfer rates are calculated implicitly to keep the donor within its Roche lobe, while post-core hydrogen burning uses the explicit Kolb scheme \citep{kolb_comparative_1990} to account for optically thin regions above the photosphere.
The mass transfer efficiency is regulated by the accretor's rotation \citep{petrovic_constraining_2005}, where material is lost from the system with the accretor's specific orbital AM once the accretor approaches critical rotation (after only a small fraction of the transferred mass is accreted, as in \cite{packet_spin-up_1981}). Orbital evolution includes the effects of mass loss and tidal interactions, assuming circular orbits.

Comparing this to the grid of \cite{temmink_coping_2023}, they differ in several aspects beyond their overshooting prescriptions.
For example, Jin et al. (in prep.) probes a higher mass range, has a finer grid and aims to follow the evolution until core carbon depletion, while the other ends the evolution after the first phase of mass transfer or at the start of core helium burning. Additionally, \cite{temmink_coping_2023} assumes nearly conservative mass transfer, while Jin et al. (in prep.) uses highly non-conservative (rotationally limited) accretion. Despite these differences, for the range of binary systems we are interested in, there is good agreement between the distributions of the stripped star masses of both binary grids.

\subsection{Analytical Expressions for the Mass transfer efficiency}

We provide an analytical approximation of the minimum mass transfer efficiency as a function of the observed sdOB and Be star masses.
It can be used to quickly estimate a lower bound on the mass transfer efficiency for Be+sdOB binaries, assuming our initial mass -- stripped mass relation is a good approximation of the true relation.

The function for our strict lower bound on $\beta$, based on the lower bound on the stripped star masses marked as (1) in Figure~\ref{fig:MinitMstrip}, is as follows (where the units of the numerical coefficients are such that all masses are in units of M$_\odot$):
\begin{equation}
  \begin{gathered}
  \beta_{\text{min}}(M_{\text{sdOB}}, M_{\text{Be}}) = \begin{cases}
      0 & \text{if } M_{\text{Be}} < M_{{1}} \\
      \frac{M_{\text{Be}} - M_{{1}}}{M_{{2}} - M_{{1}}} & \text{if } M_{{1}} \leq M_{\text{Be}} \leq M_{{2}} \\
      1 & \text{if } M_{\text{Be}} > M_{{2}}
    \end{cases} \\[0.5em]
    \begin{aligned}
      \text{where} ~
      M_{1} &= 8.52\sqrt{M_\text{sdOB} - 0.11} -1.97 + \frac{1.09}{1+M_\text{sdOB}} \\
      M_{2} &= M_{1} + 9.69(1 + 0.09M_\text{sdOB}) - \frac{10.00}{1+M_\text{sdOB}}
  \end{aligned}
  \end{gathered}
  \label{eq:beta_approximation}
\end{equation}

The function for our best guess, based on the median values of the stripped star masses marked as (2) in Figure~\ref{fig:MinitMstrip} is as follows (in units of M$_\odot$):

\begin{equation}
  \begin{gathered}
  \beta_{\text{min}}(M_{\text{sdOB}}, M_{\text{Be}}) = \begin{cases}
      0 & \text{if } M_{\text{Be}} < M_{{1}} \\
      \frac{M_{\text{Be}} - M_{{1}}}{M_{{2}} - M_{{1}}} & \text{if } M_{{1}} \leq M_{\text{Be}} \leq M_{{2}} \\
      1 & \text{if } M_{\text{Be}} > M_{{2}}
    \end{cases} \\[0.5em]
    \begin{aligned}
      \text{where} ~
      M_{1} &= 8.62\sqrt{M_\text{sdOB} - 0.13} -3.63 + \frac{2.82}{1+M_\text{sdOB}} \\
      M_{2} &= M_{1} + 9.22(1 + 0.06M_\text{sdOB}) - \frac{9.86}{1+M_\text{sdOB}}
  \end{aligned}
  \end{gathered}
  \label{eq:beta_approximation_incomplete_stripping}
\end{equation}

These approximations are accurate within $\beta$ = 0.01 for sdOB masses above 0.4 M$_\odot$, and within $\beta$ = 0.06 for even lower masses. The average error across the whole range is less than 0.001.

\section{Mass transfer efficiency correlations and initial parameters}
\label{sec:appendix_beta_constraints}
In this section we analyze correlations with other properties and infer the initial periods of the systems in our sample.

\subsection{Correlation with other properties}
\begin{figure*}[h!]
  \centering
  \includegraphics[width=\textwidth]{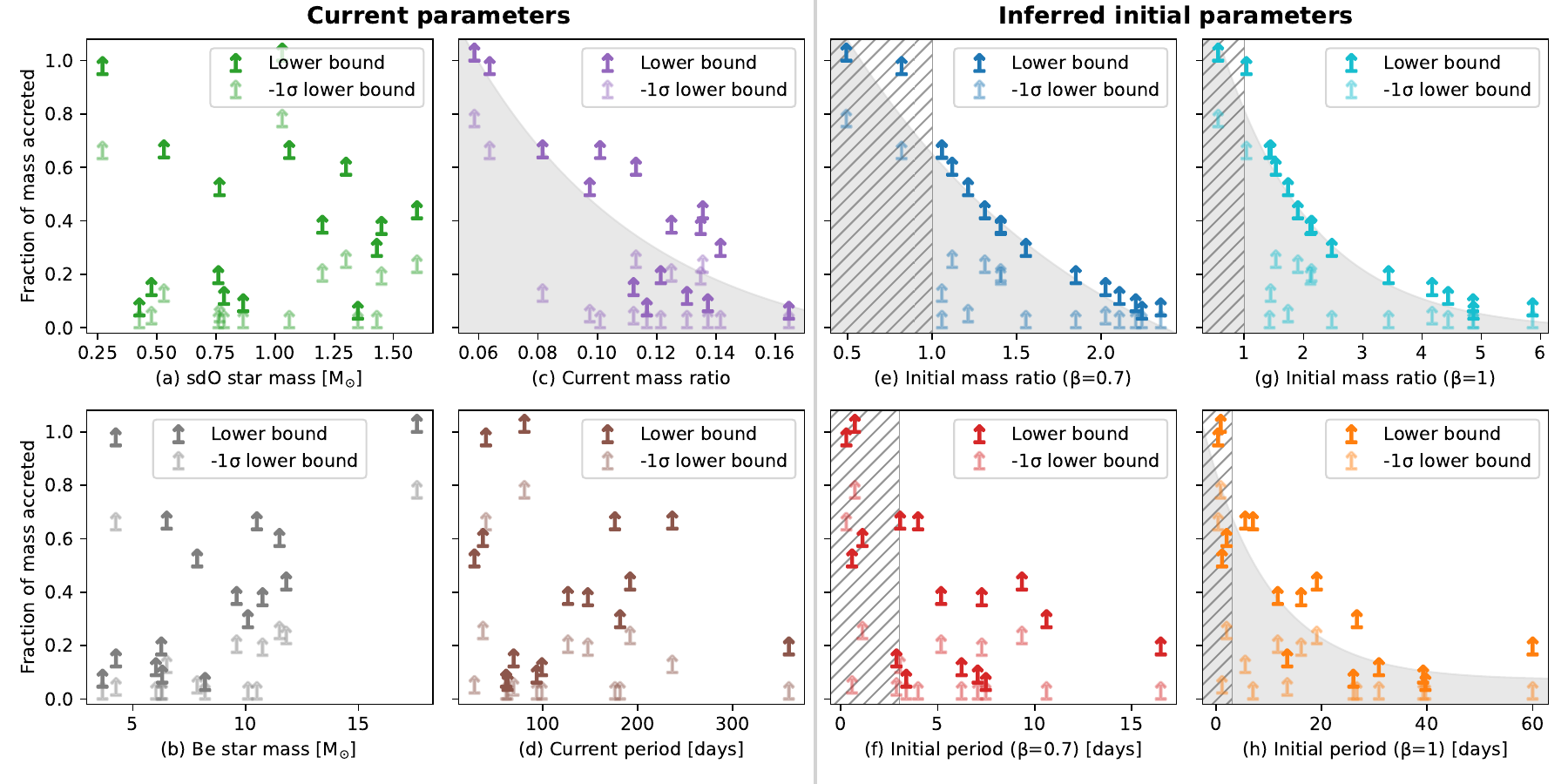}
  \caption{Correlation between the constraints on the mass transfer efficiency and current parameters (left) and inferred initial parameters (right).
  The lower bound refers to the minimum mass transfer efficiency ($\beta_{\text{min}}$) that is inferred assuming the median observed masses and the conservative stripped mass relation (Equation \ref{eq:beta_approximation}).
  The -1 sigma lower bound is based on the minimum mass transfer efficiency that is inferred within the uncertainties on the observed masses.
  The statistically significant correlations -- (c), (e), (g), (h) -- have a gray area in the background indicating their trend line and the area of the parameter space that is ruled out by our lower bounds. The hashed areas in the initial mass ratio plots indicate where the donor was initial less massive than the accretor, which is not allowed. The mass ratio is defined here as $ q = M_{\mathrm{donor}} / M_{\mathrm{acc}}.$
  The hashed areas in the initial period plots indicate Case A mass transfer.
   }
  \label{fig:correlation}
\end{figure*}

In Figure~\ref{fig:correlation} we investigate the correlation between the mass transfer efficiency constraints and other properties of the systems, such as the observed masses, the mass ratios, and the periods of the systems.
For the latter two, we show both the correlation with the current state of the system, as well as the correlation with the initial state of the system.
For the initial state, we need to assume a mass transfer efficiency, which we show for $ \beta = 0.7 $ and $ \beta = 1 $.
For the initial period reconstruction, we assume isotropic re-emission of the fraction of the transferred mass that was not accreted by the Be star, which is explained in more detail in the next section.
The mass ratio is defined here as $ q = M_{\mathrm{donor}} / M_{\mathrm{acc}} $.

There is a clear trend of an increasing lower bound on $ \beta $ for mass ratios closer to unity.
This is expected, because the initial mass ratio constraint, which says the Be star must be less massive than the sdOB star before mass transfer, is directly tied to the mass transfer efficiency constraint.
Another, less strong trend is that the lower bound on $ \beta $ increases with decreasing initial period.

In order to statistically test and validate these correlations, we calculate the Kendall rank correlation coefficient \citep{kendall_new_1938} for each property with the mass transfer efficiency constraints.
This is a non-parametric test that measures the statistical dependence between variables without assuming any particular distribution. It is similar to the Spearman's rank correlation \citep{spearman_proof_1904}, but is considered better suited for small datasets due to being more robust to ties and outliers.
This makes it well-suited for our sample of 16 systems. The results of this test confirm the visual trends.
The correlation with initial mass ratio is strong and statistically significant.
The correlation with the current mass ratio and initial periods is only moderate but still significant (depending on the assumed $\beta$). The correlation with other properties is not statistically significant.

Both the initial mass ratio and initial period correlations can be consistent with theoretical expectations.
For example, at extreme initial mass ratios or long initial periods, the thermal timescales of both stars are very different at the time of Roche lobe overflow, making it unlikely for mass transfer to occur in a stable fashion. Additionally, the AM transfer could be lower in closer systems, allowing the accretor to retain more mass.
While this qualitatively fits the theoretical predictions, they differ quantitatively from our allowed $\beta$ ranges, as shown in the main text.

Note that we do not find a correlation of $\beta$ with initial donor mass in our mass range of $2 - 9$  M$_\odot$, since panel (a) shows no correlation with the stripped star mass and the initial mass is related 1:1 with the stripped star mass using Figure~\ref{fig:MinitMstrip}.

\subsection{Initial periods}
\begin{figure*}[h]
  \centering
  \includegraphics[width=\textwidth]{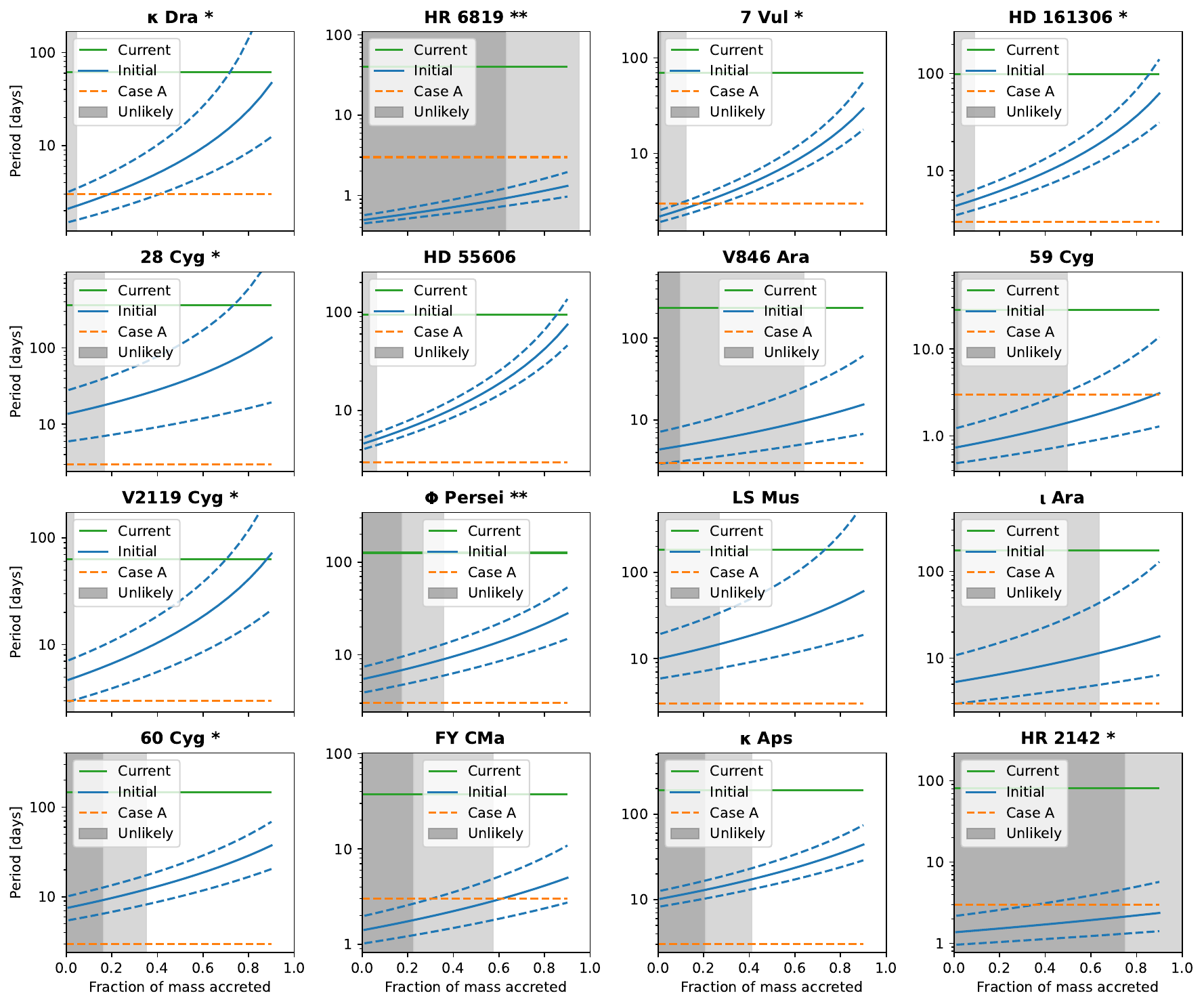}
  \caption{
    \textbf{The inferred initial periods for each system in our sample.}
  The solid green horizontal line shows the current period, the blue curved line shows the initial period. The blue dashed lines show the 1 sigma uncertainty of the observed masses and their effect on the initial period.
  We assume that 10\% of the transferred mass is sent to a circumbinary disk, and that the mass that is not kept by the accretor is re-emitted isotropically.
  To infer the initial mass ratio, we use the conservative stripped mass relation (Equation \ref{eq:beta_approximation}).
  The orange dashed line shows the initial period below which Case A mass transfer would occur (roughly).
  The grayed out areas show our inferred minimum mass transfer efficiencies using our strict estimate (Equation \ref{eq:beta_approximation}), assuming the median observed masses (light gray) or the lowest minimum mass transfer efficiency within 1 sigma of the median observed masses (dark gray).
  Mass transfer efficiencies that correspond to initial periods that are above the Case A boundary, are consistent with Case B mass transfer.
  The asterisks next to the names of the systems show the characterization tiers, as in Table~\ref{table:sample1}.
  }
  \label{fig:period_constraints_beta_disk_0.1}
\end{figure*}

In order to check the consistency of our results, we inferred the initial periods of the systems using equation 4.58 from \cite{tauris_physics_2022} \citep[see also][]{huang_modes_1963,soberman_stability_1997}{}
We consider the case where a part (i.e. 10\%) of the transferred mass is sent to a circumbinary disk with radius $2 \times a$, and assume isotropic re-emission of the fraction of the transferred mass that was not accreted by the Be star.
Constraints on the amount of angular momentum lost during mass transfer are rare, but we estimate that this is a good approximation for the majority of our sample.
The inferred initial periods are shown in Figure~\ref{fig:period_constraints_beta_disk_0.1}. For almost all systems, we find initial periods that are consistent with Case B mass transfer.
The exceptions are 59 Cyg, which is only marginally consistent, HR 2142 and HR 6819.
The latter two are notable outliers, as discussed in the main text. These systems require a larger amount of mass to be sent to a circumbinary disk (above 20\%) to be consistent with Case B, and together with the nearly completely conservative mass transfer efficiency constraints, are more likely to have gone through Case A mass transfer.

\section{Comparison with thermally limited accretion}\label{sec:appendix_thermally_limited}
\begin{figure*}[h]
  \centering
  \begin{subfigure}[b]{0.6\textwidth} %
      \centering
      \includegraphics[trim=10pt 12pt 10pt 10pt, clip, width=\linewidth]{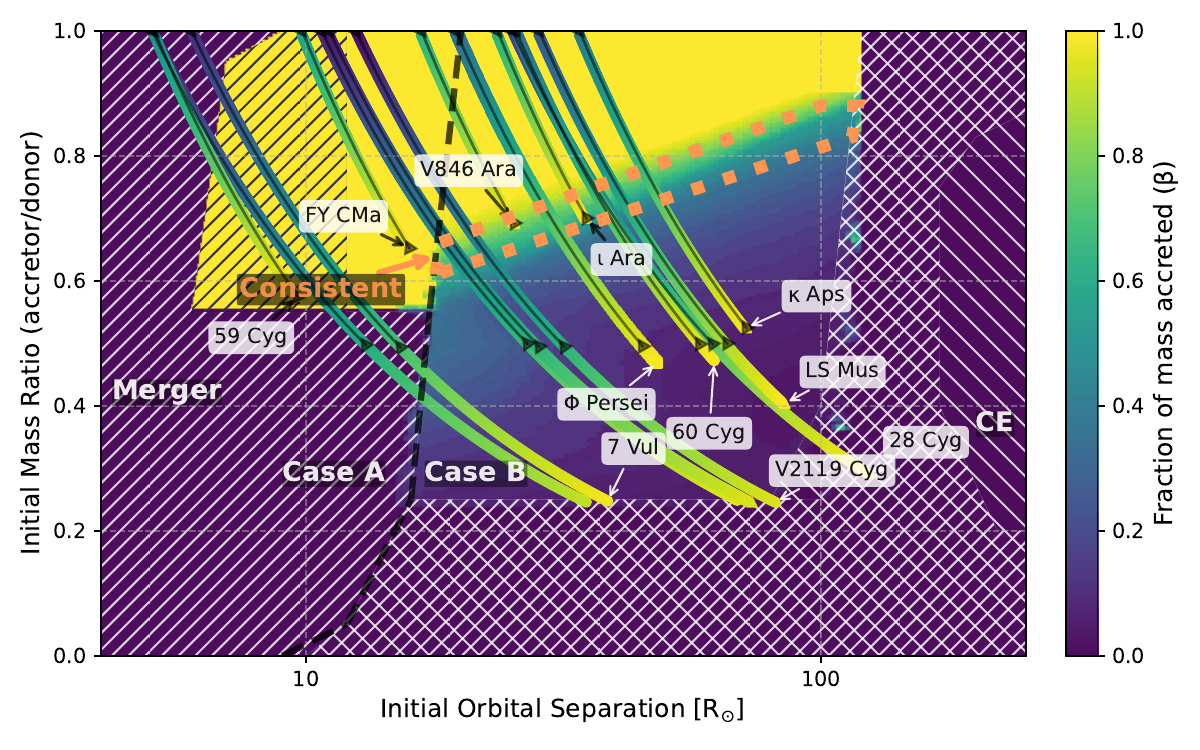}
      \caption{Isotropic re-emission.}
      \label{fig:schneider_complete_0}
  \end{subfigure}

  \begin{subfigure}[b]{0.6\textwidth} %
      \centering
      \includegraphics[trim=10pt 12pt 10pt 10pt, clip, width=\linewidth]{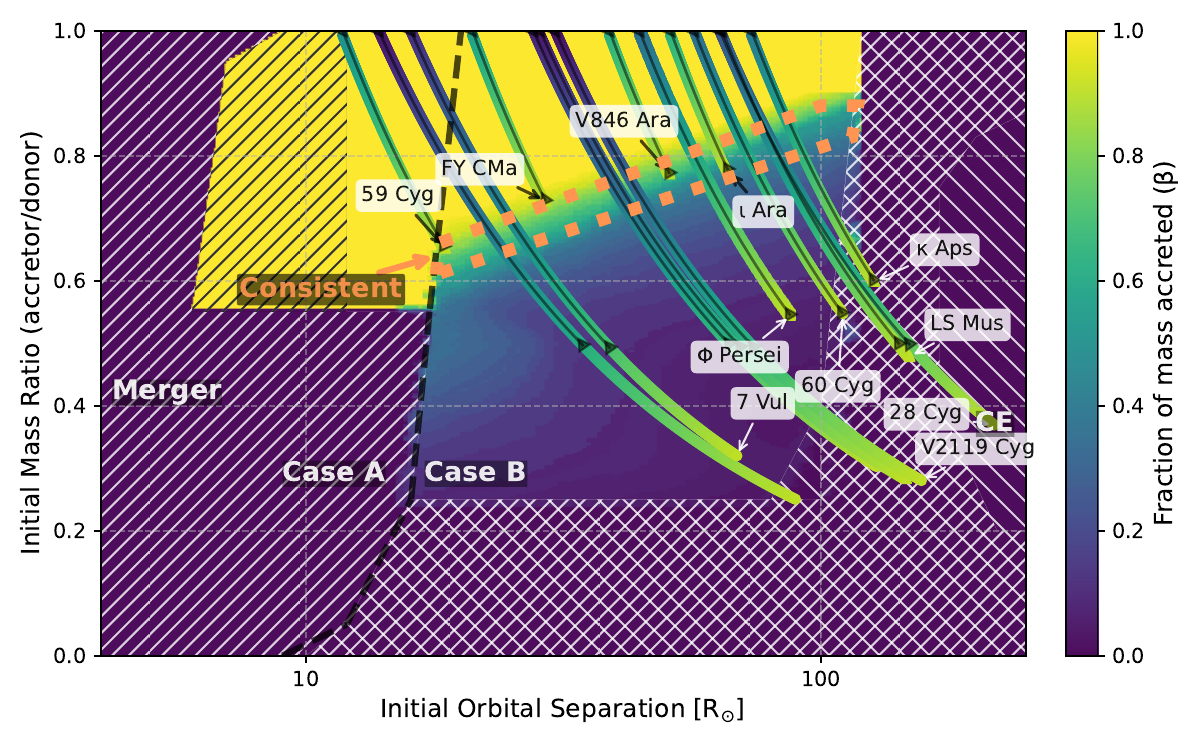}
      \caption{Assuming 10\% of the transferred mass is sent to a circumbinary disk.}
      \label{fig:schneider_complete_10}
  \end{subfigure}

  \begin{subfigure}[b]{0.6\textwidth} %
      \centering
      \includegraphics[trim=10pt 12pt 10pt 10pt, clip, width=\linewidth]{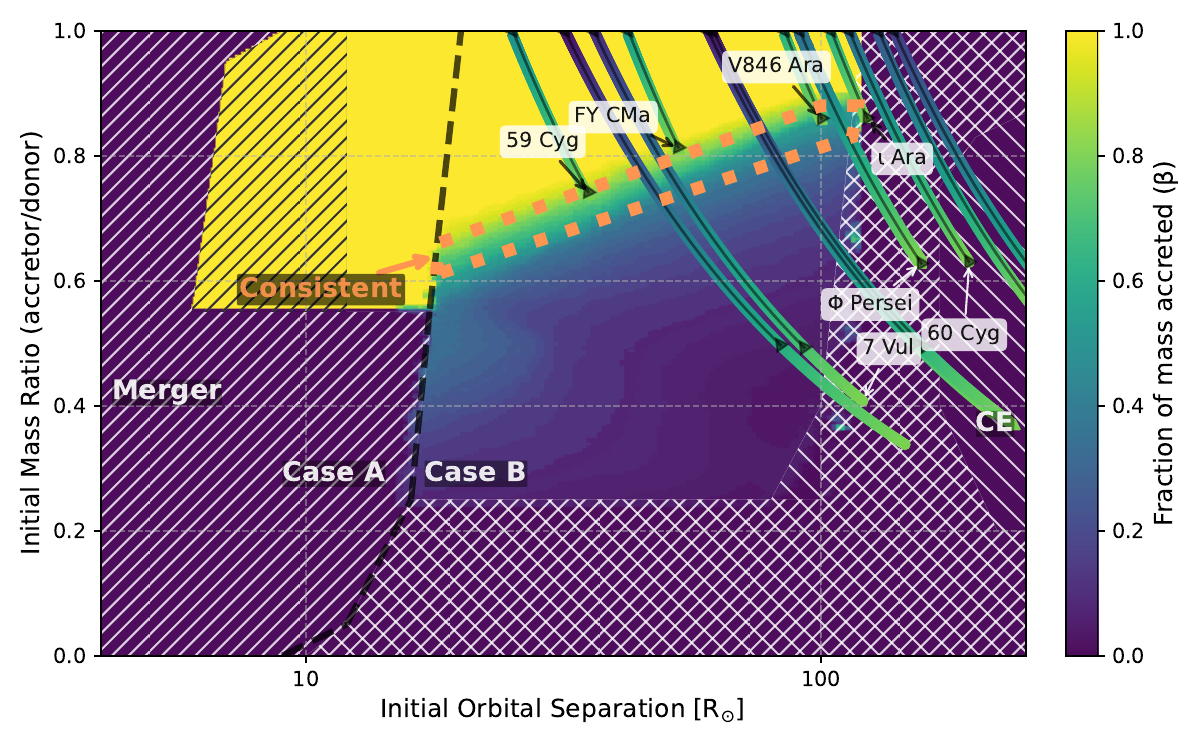}
      \caption{Assuming 20\% of the transferred mass is sent to a circumbinary disk.}
      \label{fig:schneider_complete_20}
  \end{subfigure}

  \caption{Like Figure~\ref{fig:schneider_incomplete_10}, but assuming complete stripping of the envelope (using Equation \ref{eq:init-strip_conservative}) and varying the assumptions of angular momentum loss: (a) assuming isotropic re-emission of the non-accreted mass, (b) assuming 10\% of the transferred mass is sent to a circumbinary disk and the rest of the non-accreted mass is re-emitted isotropically, (c) assuming 20\% of the transferred mass is sent to a circumbinary disk and the rest of the non-accreted mass is re-emitted isotropically.}

  \label{fig:schneider_complete_comparison} %
\end{figure*}

We compared the consistency of the thermally limited accretion model in \cite{schneider_evolution_2015} with our observed sample in Figure~\ref{fig:schneider_incomplete_10} in the main text.
In that model, the Case A/Case B boundary is at an initial orbital separation of roughly 15-20 R$_\odot$. The hashing indicates initial parameters that likely lead to mergers or common envelope (CE) scenarios.
On top of the background, which shows the predictions of the thermally limited accretion model, we plot the inferred initial parameters of our observed sample, which depend in part on the assumed $\beta$.
Figure~\ref{fig:schneider_incomplete_10} in the main text showed that the main area where the predictions of the thermally limited accretion model match the observations is quite narrow, indicating tension with the model. However, it was only shown for one set of assumptions, namely that there was incomplete stripping of the envelope and that 10\% of the transferred mass was sent to a circumbinary disk.
In Figures \ref{fig:schneider_complete_0}, \ref{fig:schneider_complete_10}, and \ref{fig:schneider_complete_20}, we show three alternative assumptions. Namely, we now assume complete stripping of the envelope and that 0\%, 10\% or 20\% of the transferred mass is sent to a circumbinary disk, respectively.
Although the locations of the systems change considerably, the area where the $\beta$'s match remains roughly constant. This is due to the steep predicted $\beta$ gradient of the thermally limited accretion model, which only allows for partially conservative accretion (with $\beta = 0.3 - 0.9$) in a limited range.

\end{appendix}

\bibliography{bibliography}{}
\bibliographystyle{aasjournalv7}

\end{document}